\documentclass[10pt,aps,pra,showpacs,floatfix,floats,superscriptaddress,twocolumn]{revtex4-1}

\usepackage{graphicx}
\usepackage{latexsym}
\usepackage{amssymb,amsmath,bm}
\usepackage{xcolor}
\usepackage{xparse}
\usepackage{etoolbox}
\NewDocumentCommand \BiG {m O{} O{}}{                                                                                                                                 
         \IfNoValueTF{ #2}                                                                                                                                             
         {                                                                                                                                                             
                 \IfNoValueTF{ #3}                                                                                                                                     
                 {                                                                                                                                                     
                         \big{.#1}                                                                                                                                     
                 }                                                                                                                                                     
                 {                                                                                                                                                     
                         \big{.#1}^{ #3 }                                                                                                                              
                 }                                                                                                                                                     
         }                                                                                                                                                             
         {                                                                                                                                                             
                 \IfNoValueTF{ #3}                                                                                                                                     
                 {                                                                                                                                                     
                                                                                                                                                                       
                         \big{.#1}_{ #2}                                                                                                                               
                 }                                                                                                                                                     
                 {                                                                                                                                                     
                         \big{.#1}_{ #2}^{ #3}                                                                                                                         
                 }                                                                                                                                                     
         }                                                                                                                                                             
}
\DeclareMathOperator{\vctrz}{vec} 
\DeclareMathOperator{\Diag}{Diag} 
\newcommand{\kronprod}[2]{        
        \ifstrequal{#1}{\mathcal{I}}{\left(#1 \otimes #2 \right)}{\left(#1^T \otimes #2 \right)}                                                                      
} 

\begin{document}
\title{Regimes of radiative and nonradiative transitions in transport through\\ an electronic system
       in a photon cavity reaching a steady state}

\author{Vidar Gudmundsson}
\email{vidar@hi.is}
\affiliation{Science Institute, University of Iceland, Dunhaga 3, IS-107 Reykjavik, Iceland}
\author{Thorsteinn H.\ Jonsson}
\affiliation{Science Institute, University of Iceland, Dunhaga 3, IS-107 Reykjavik, Iceland}
\author{Maria Laura Bernodusson}
\affiliation{Physikalisches Institut, Albert-Ludwigs-Universit{\"a}t Freiburg,
             Hermann-Herder-Str.\ 3, D-79104 Freiburg, Germany}
\affiliation{Science Institute, University of Iceland, Dunhaga 3, IS-107 Reykjavik, Iceland}
\author{Nzar Rauf Abdullah}
\affiliation{Physics Department, College of Science, 
             University of Sulaimani, Kurdistan Region, Iraq}
\author{Anna Sitek}
\affiliation{Science Institute, University of Iceland, Dunhaga 3, IS-107 Reykjavik, Iceland}
\affiliation{Department of Theoretical Physics, Faculty of Fundamental Problems of Technology, 
             Wroc{\l}aw University of Technology, 50-370 Wroc{\l}aw, Poland}
\author{Hsi-Sheng Goan}
\email{goan@phys.ntu.edu.tw}
\affiliation{Department of Physics and Center for Theoretical Sciences, National Taiwan University, 
             Taipei 10617, Taiwan}
\affiliation{Center for Quantum Science and Engineering, 
             National Taiwan University, Taipei 10617, Taiwan}
\author{Chi-Shung Tang}
\email{cstang@nuu.edu.tw}
\affiliation{Department of Mechanical Engineering, National United University, Miaoli 36003, Taiwan}
\author{Andrei Manolescu}
\email{manoles@ru.is}
\affiliation{School of Science and Engineering, Reykjavik University, Menntavegur 
             1, IS-101 Reykjavik, Iceland}

%

\begin{abstract}
We analyze how a multilevel many-electron system in a photon cavity
approaches the steady state when coupled to external leads. 
When a plunger gate is used to lower cavity photon dressed one- and two-electron
states below the bias window defined by the external leads, we 
can identify one regime with nonradiative transitions dominating 
the electron transport, and another regime with radiative 
transitions. Both transitions trap the electrons in the states below the bias
bringing the system into a steady state. The order of the two
regimes and their relative strength depends on the location of the bias window in the energy
spectrum of the system and the initial conditions.
\end{abstract}

\maketitle

\section{Introduction}
Quantum master equations have been introduced to study transient or steady 
states in open quantum systems coupled to their environment at some 
point in time \cite{Zwanzig60:1338,Nakajima58:948,Haake73:98}.
Among many other systems this approach has been used for cavity-QED 
systems \cite{PhysRevB.81.155303,PhysRevB.87.195427}, as well as systems where 
electrons are transported through a photon 
cavity \cite{doi:10.1021/acsphotonics.5b00115}.
Commonly, the interest of researchers has been on the steady 
state \cite{PhysRevE.91.013307} of a two-level system with a Markovian master 
equation in a Lindblad form \cite{Lindblad76:119}.
In light of experiments on electrons transport through
photon cavities \cite{PhysRevLett.113.036801,Peterson380:2012,PhysRevX.6.021014},
the transient regime has been investigated 
for more complex electronic systems with
non-Markovian master equations for a weak coupling to the 
leads \cite{PhysRevB.87.035314,0953-8984-25-46-465302}. 

When the lowest electron states are within the bias window the steady
state can be reached within a time that is accessible by direct time-integration
of the non-Markovian master equation in the weak coupling limit for a truncated 
Fock space with 120 many-body states \cite{0953-8984-25-46-465302}. For the case
of few or several dressed electron states below the bias window this is not possible
any more and one must keep in mind that nonradiative and radiative relaxation 
channels with characteristic time constants differing by several orders of magnitude
are at work. 

In the steady state limit the master equation turns into an algebraic 
equation that can conveniently be solved in 
Liouville space 
\cite{Weidlich71:325,PhysRevB.81.155303,Nakano2010,PhysRevA.87.012108,0953-8984-24-22-225304,HarbolaPR465:2008,1367-2630-8-2-021,PhysRevB.84.134501}.
In this letter we show that this approach can be used to attain the steady
state information and the Markovian time-evolution over many orders of magnitude for the 
time variable for a complex electron system in a photon cavity coupled both to external 
leads and a photon reservoir. 

\section{Model} 
\begin{figure}[htb]
      \includegraphics[width=0.42\textwidth]{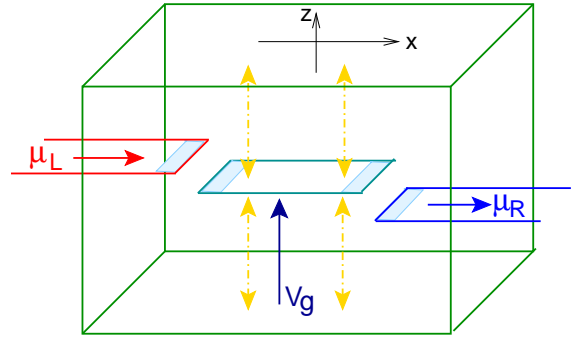}
      \caption{A sketch of the central electronic system in a rectangular photon cavity 
               coupled to external semi-infinite leads with chemical potentials
               $\mu_L$ and $\mu_R$. The leads are kept at temperature $0.5$ K, and
               the contact area is indicated with light blue shading. The action of a plunger gate
               is marked by $V_g$ and the golden arrows represent the cavity photons.}
      \label{Sketch}
\end{figure}

The central system is a short GaAs quantum wire schematically shown in Fig.\ \ref{Sketch} 
with parabolic confinement with characteristic energy $\hbar\Omega_0=2.0$ meV along the $y$-direction
and length $L_x=150$ nm along the transport direction, $x$. 
An external magnetic field $B=0.1$ T perpendicular to the $x-y$-plane is used to break
the spin degeneracy, and leads to a natural length scale $a_w=(\hbar /(m^*\Omega_w))^{1/2}$,
where $\Omega_w=(\Omega_0^2+\omega_c^2)^{1/2}$, and $\omega_c=(eB/m^*c)$ is the cyclotron
frequency. The electrons interact mutually via the Coulomb interaction and with the 
photons of a single mode in a rectangular cavity with both the para- and diamagnetic 
electron-photon interactions. The interactions are treated stepwise with exact numerical 
diagonalization in truncated Fock spaces assuming the size of the cavity much larger 
than the extent of the electron system \cite{Gudmundsson:2013.305,Gudmundsson16:AdP}. 
The leads are semi-infinite in the same external perpendicular magnetic field as the 
central system, and with the same parabolic lateral confinement. Before the coupling to the central 
system the electrons in each lead are viewed as a non-interacting Fermi gas with a 
chemical potential $\mu_l$ ($l=L,R$) describing their equilibrium state. The leads
end in a hard wall confinement as does the central system and their coupling Hamiltonian 
is constructed using a nonlocal overlap function of the single-electron states on either
side of the hard wall interface. 
The coupling of the leads and the central system does thus depend
on the shape of the wavefunctions in the `contact area' (see Fig.\ \ref{Sketch}) and also on the electron `affinity' of
the states in the leads and the central system through a factor $\exp{(-|\epsilon_{ql}-E_i|/\Delta_E^l)}$,
where $\Delta_E^l=0.5$ meV, $\epsilon_{ql}$ is the energy of a single-electron state
in the leads, and $E_i$ the corresponding energy eigenvalue in the central system 
\cite{Gudmundsson09:113007,Gudmundsson16:AdP}.

The time-evolution of the reduced density operator of the central system is described by
a non-Markovian integro-differential equation \cite{Moldoveanu10:155442}, not in a Lindblad form 
\begin{equation}
       \partial_t \rho = -\dfrac{i}{\hbar} \left[H,\rho \right] -  \sum_{l=\{L,R\}}\Lambda(\Omega_{ql},\tau_{ql},t), 
\label{GME}
\end{equation}
with the dissipation term
\begin{equation}
      \Lambda(\Omega_{ql},\tau_{ql},t) = \dfrac{1}{\hbar^2} \int dq\; \left\lbrace
          \left[\tau_{ql},{\Omega_{ql}}(t) \right] + h.c. \right\rbrace, 
\label{Dissipation}
\end{equation}
where the time integral is expressed by,   
\begin{align}
	\Omega_{ql}(t) =  \int_{0}^{t} ds\; & U(t-s) \left\{ \tau^\dag_{ql} {\rho}(s) (1 - f_{ql})\right.\nonumber\\
                           &\left. - {\rho}(s) \tau^\dagger_{ql} f_{ql} \right\} U^{\dagger}(t-s)e^{i(s-t)\omega_{ql}}.
\label{Omega}
\end{align}
This result is derived by applying the Nakajima-Zwanzig formalism \cite{Zwanzig60:1338,Nakajima58:948}, projecting 
the Liouville-von Neumann equation for the density operator of the total system onto the central system assuming
only second order terms in the coupling in the dissipative term.
The $q$ integration is over the `momentum' variable in the leads and a summation over their
subband indices. The coupling of the subsystems, described by the coupling tensor $\tau_{ql}$, is suddenly
turned on at $t=0$. The energy spectrum of lead $l$ is represented by
$\omega_{ql}=\epsilon_{ql}/\hbar$, and $U$ is the time evolution operator for the closed central system. 
The Fermi function describing the equilibrium in each lead before the coupling is $f_{ql}$.

In order to search for the steady state of the system and eventually the time-evolution for large times we
perform a Markov approximation, but avoiding any other such as the rotating wave approximation. For matrix elements of 
Eq.\ (\ref{Omega}) after the change of variable $t-s\rightarrow s'$ we set $\rho (t-s)\rightarrow \rho (t)$ and use  
$\int_0^tds \exp{[is(E_\nu -E_\mu -\epsilon_{ql})]}\rightarrow \pi\delta(E_\nu -E_\mu -\epsilon_{ql})$.
The Dirac delta function is used to transform Eq.\ (\ref{Dissipation}) by noting that for
 any operator $A$ in the central system, 
$\int dq {A}(q)\delta(E_\beta -E_\alpha -\epsilon_{ql})=
\int d\epsilon (dq/d\epsilon )A(\epsilon )\delta(E_\beta -E_\alpha -\epsilon)= 
{A}^{\beta \alpha}D^{\beta \alpha}$, where $D$ is the density
of states in the leads and the upper indices denote the value of the operator at $E_\beta - E_\alpha$. Thus yielding a
reprised form of Eq.\ (\ref{Omega}),
\begin{equation}
      \BiG{\Omega}[\alpha \beta] =  \left\lbrace \BiG{\mathcal{R}[\rho]}[\alpha \beta] -
          \BiG{\mathcal{S}[\rho]}[\alpha \beta]\right\rbrace \delta^{\beta \alpha},
\label{OmegaMat}
\end{equation}
where we have defined $\mathcal{R}[\rho] = \rho \pi f \BiG{\tau}[][\dagger]$, and $\mathcal{S}[\rho] = \pi (1-f)\BiG{\tau}[][\dagger]\rho$.
The upper indices in (\ref{OmegaMat}) refer to a `Bohr frequency' in the energy spectrum of the central system
and the lower ones to a matrix element. 
The steady state is found from Eq.\ (\ref{GME}) with $\partial_t\rho =0$. 
In order to devise an efficient parallel computational scheme of the algebraic equation resulting in the combining of
Eqs.\ (\ref{OmegaMat}) and (\ref{Dissipation}), we resort to map the Markovian master equation for the steady state to
Liouville space using vectorization of matrices and Kronecker tensor products. 
%

%
Any possible configuration of components from the dissipation term of the 
Markovian master equation is encapsulated by the following equation
\begin{equation} \label{eq:generalcomprep}                                                                                                                                                     
        \BiG{Z}[\alpha \beta] = \int \BiG{D}[][]\BiG{A}[\alpha \lambda][] \BiG{\Omega}[\lambda \sigma][] \BiG{B}[\sigma \beta][]d\BiG{\delta}[][\sigma\lambda] ,
\end{equation}
where $Z, A, B$ act as placeholders for any operators in the Fock space, 
other operators being defined as before. 
It can be observed from Eq.\ (\ref{OmegaMat}) and (\ref{eq:generalcomprep}) that the $\Omega$ 
operator induces a Dirac delta measure for each of its
components, related to the components of $\mathcal{S}$ and $\mathcal{R}$. 
Through this relation any component of $Z$ reveals the amount of dissipation in the transitions constituting
the Liouville space. Seeking a matrix representation of such an equation we construct a matrix 
of Dirac delta measure such that
\begin{equation}
      \Delta_{\alpha \beta} = \delta^{\alpha \beta} = \delta{(E_\alpha -E_\beta -\epsilon )}.
\end{equation}
We will call this the Dirac matrix.
Using the Dirac matrix we obtain a matrix representation of Eq.\ (\ref{eq:generalcomprep}) by applying the 
Hadamard product of its transpose to the factors adhering to $\Omega$
\begin{equation}                                                                                                                                                      
         Z = \int D A\left\lbrace \left( \mathcal{R}[\rho] - \mathcal{S}[\rho] \right) \odot d\Delta^T \right\rbrace B.
\end{equation}
In general the Hadamard product is a non-linear transformation of matrices which implies that the diligent 
process of working with the component-representation could be ineffective unless numerical methods 
are pursued as in \cite{PhysRevE.91.013307}. 
Having the matrix representation of the equation allows us to map it to Liouville space. 
This is done by applying the vectorization operator, 
obtaining the following form 
\begin{align}                                                                                                                                              
 \vctrz(Z) =& \int \kronprod{B}{DA}\vctrz\left(\left\lbrace\mathcal{R}[\rho] -
             \mathcal{S}[\rho]\right\rbrace \odot d\Delta^T \right) \nonumber\\
             &\hspace*{0.4cm}+ (h.c.)_\mathrm{vec} ,      
\end{align}
and subsequently using the identity
\begin{equation}  
      \vctrz\left( AB \odot \Delta^T \right) = 
      \Diag(\Delta^T)\left(B^T\otimes\mathcal{I}\right)\vctrz(A).
\end{equation}
This leads the dissipation term from Eq.\ (\ref{Dissipation}) to be viewed as a linear operator acting on the density operator in Liouville Space. That is
\begin{equation}
        \vctrz(\Lambda[\rho]) = \sum_{X = R,S}\left(\mathfrak{Z}_{X_1}\mathfrak{Z}_{X_2}\right)\vctrz(\rho) .
\label{vecLr}
\end{equation}
The matrices of the linear maps $\mathfrak{Z}_{R_1},
\mathfrak{Z}_{R_2}, \mathfrak{Z}_{S_1}, \mathfrak{Z}_{S_2} $ are given by
\begin{align}
        \mathfrak{Z}_{X_1} &= \int \kronprod{{B}}{D{A}}\Diag(d\Delta^T) ,\qquad X = R,S ,\\
        \mathfrak{Z}_{R_2} &= \quad\int \Diag(d\Delta^T)({I}\otimes{R}) ,\\
        \mathfrak{Z}_{S_2} &= -\int \Diag(d\Delta^T)\kronprod{S}{I} ,
\end{align} 
where $R = \mathcal{R}[I]$ and $S = \mathcal{S}[I]$ with $I$ being the identity operator.
These matrices can be constructed in an efficient 
manner given the proper computational resources. 

If we let, $\partial_t\rho_{\mathrm{vec}} = \mathfrak{L}\rho_{\mathrm{vec}}$, stand for the Markovian master equation in the
Liouville space then we can find the left and right eigenvectors satisfying,
$\mathfrak{L}\mathcal{V}=\mathcal{V}\mathfrak{L}_{\mathrm{diag}}$, and 
$\mathcal{U}\mathfrak{L}=\mathfrak{L}_{\mathrm{diag}}{\mathcal{U}}$, with $\mathcal{UV}=\mathcal{VU}=\mathfrak{1}$, and the time evolution
is \cite{PhysRevB.81.155303}
\begin{equation}
      \rho_{\mathrm{vec}} (t) = \left[\mathcal{U}\exp{(\mathfrak{L}_\mathrm{diag}t)\mathcal{V}} \right]\rho_{\mathrm{vec}} (0).
\label{t-evolution}
\end{equation}
For a complex system with large Fock and Liouville spaces the proper limit of this equation is a reliant
way to search the steady state. 

\section{Results} 
\begin{figure}[htb]
      \includegraphics[width=0.48\textwidth]{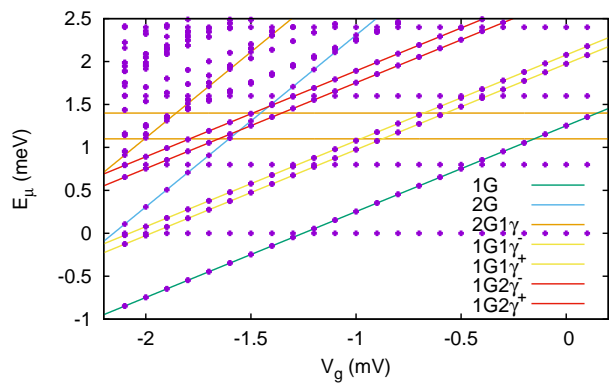}
      \caption{The many-body energy spectrum of the closed central system as
               a function of the plunger gate voltage $V_g$. The golden horizontal lines indicate the
               chemical potential of the left lead $\mu_L=1.4$ meV, and the right lead $\mu_R=1.1$ meV.
               1G denotes the one-electron groundstate, 2G the two-electron one, 1G1$\gamma^\pm$
               stands for the Rabi-split first replica of the 1G, and 1G2$\gamma^\pm$ for the 
               second split replica. $B=0.1$ T, $g_{\mathrm{EM}}=0.05$ meV,
               $m^*=0.067m_e$, and the effective Land{\'e} g-factor $g=-0.44$.}
      \label{Fig-Erof}
\end{figure}

\begin{figure}[htb]
      \includegraphics[width=0.48\textwidth]{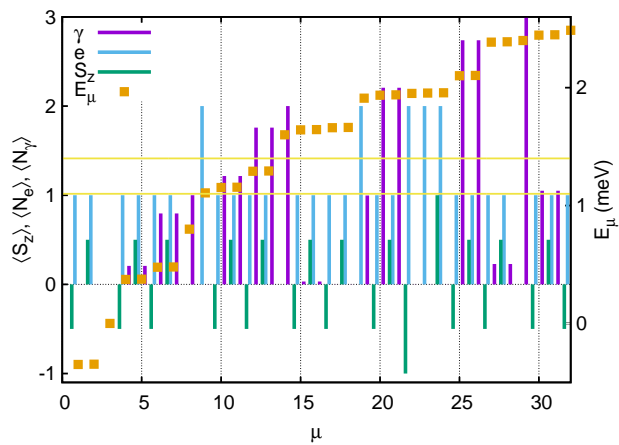}
      \caption{The energy $E_\mu$, the electron number $N_e$, the mean photon number $N_\gamma$,
               and the spin $s_z$ for each many-body state $|\mu )$ for $V_g=-1.6$ mV.
               The yellow horizontal lines indicate $\mu_L$ and $\mu_R$. The vacuum state is
               the third state.
               Other parameters as in Fig.\ \ref{Fig-Erof}.}
      \label{NeNphESz-Vgm1p6-32}
\end{figure}

A low energy section of the spectrum of the fully interacting many-body states, $|\mu )$, 
of the closed central system is 
displayed in Fig.\ \ref{Fig-Erof} as a function of the plunger gate voltage $V_g$.
Without the cavity photon field the energy difference between the one-electron groundstate
and the first excitation thereof is approximately 0.744 meV. The single-mode $x$-polarized 
{strongly coupled} photon field with energy 0.8 meV causes a Rabi-splitting seen in the spectrum.
The properties of the 32 lowest dressed electron and photon states is analyzed in 
Fig.\ \ref{NeNphESz-Vgm1p6-32} for $V_g=-1.6$ mV, and $g_{\mathrm{EM}}=0.05$ meV, 
{the electron-photon coupling strength}.
A noninteger photon number characterizes the Rabi-split dressed electron states. 
{The values for the plunger gate
voltage and the photon energy are selected here in order to have the bias window rather high in the 
energy spectrum to give internal processes in the short quantum wire weight on the path to a steady state.}

\begin{figure}[htb]
      \includegraphics[width=0.48\textwidth]{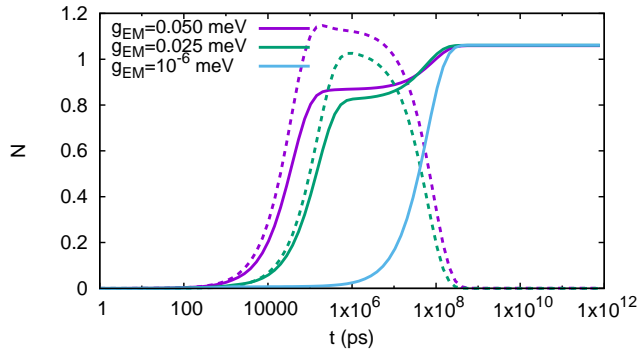}
      \caption{The mean values of the total electron (bold) and photon (dotted) numbers
               as functions of time for different values of the electron-photon coupling $g_\mathrm{EM}$
               for $V_g=-1.6$ mV.}
      \label{Fig-NeNg}
\end{figure}

\begin{figure}[htb]
      \includegraphics[width=0.48\textwidth]{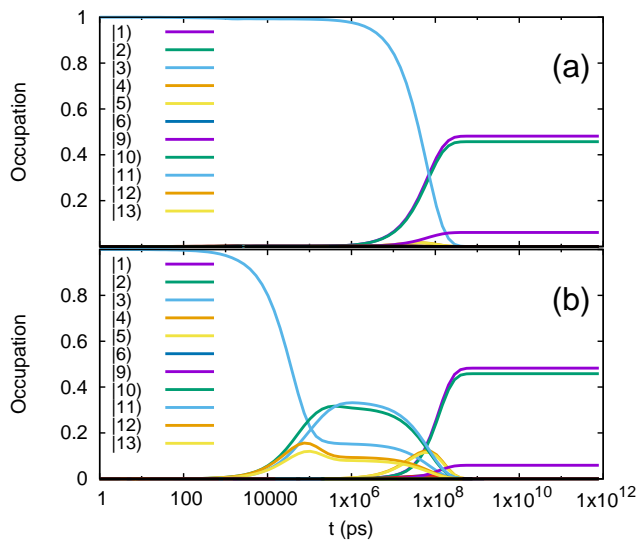}
      \caption{The mean occupation of the many-body states as a function of time, for
               (a) $g_\mathrm{EM}=1\times 10^{-6}$ meV, and (b) $g_\mathrm{EM}=0.05$ meV for $V_g=-1.6$ mV.
               Only states with significant occupation are listed.}
      \label{Fig-occ}
\end{figure}

For $V_g=-1.6$ mV and the time $t=0$ the leads and the central system in the vacuum eigenstate $|3)$ 
are abruptly, {but weakly coupled} with an overall 
coupling constant $g_{\mathrm{LR}}a_w^{3/2}=0.124$ meV, and the Markovian time-evolution according to Eq.\ (\ref{t-evolution})
is shown in Fig.\ \ref{Fig-NeNg} for the total mean electron (bold) and photon (dotted) number for three values of the 
photon-electron coupling $g_\mathrm{EM}$. 
For all cases the system has reached a steady state for time $t>10^{-3}$ s, as is confirmed in Fig.\ \ref{Fig-occ}
showing the steady state to be mainly composed of the two spin components of the one-electron ground state,
$|1)$ and $|2)$, and 
with a small contribution of the two-electron groundstate $|9)$. In the bias window are 5 states, the two-electron
groundstate $|9)$ with vanishing photon component, both spin components of the lower Rabi branch of the one-electron groundstate,
$|10)$ and $|11)$ with approximately 0.02 0-photon, 0.75 1-photon and 0.23 2-photon contribution, and the upper Rabi branch,
$|12)$ and $|13)$ with approximately 0.01 0-photon, 0.22 1-photon and 0.77 2-photon contribution. As Fig.\ \ref{Fig-occ}(a)
shows the coupling of the leads states is very small to the groundstate and the system needs a long time to be charged
with vanishing electron-photon coupling. On the other hand Fig.\ \ref{Fig-NeNg} shows a faster charging for finite 
electron-photon coupling and Fig.\ \ref{Fig-occ}(b) demonstrates that in this case charging occurs through the Rabi split states,
$|10)$-$|13)$ in the bias window initially, with some {charge} reappearing in lower lying Rabi-split states, 
$|4)$ and $|5)$, with mean photon content 0.2, before ending in the two spin components of the one-electron groundstate, $|1)$ and $|2)$. 
The mean photon number in Fig.\ \ref{Fig-NeNg} confirms this contribution of states with a photon component to the transport,
albeit at a much later time than a resonant tunneling through a state without a photon component would need, for example the 
ground state for $V_g=0$ with the present bias window needs 100 ps to get considerable charge and will reach the steady state
for $t>1$ ns. Not seen very clearly on the logarithmic time scale is the fact that the systems reaches the steady state a bit
later in the case of an electron-photon coupling.   
\section{Discussion} 
\begin{figure}[htb]
      \includegraphics[width=0.48\textwidth]{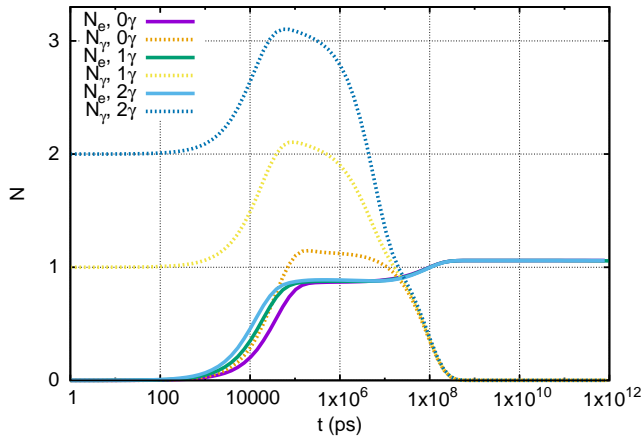}
      \caption{{The mean values of the total electron (bold) and photon (dotted) numbers
               as functions of time for different values of photons initially present in the cavity
               at $t=0$ for $V_g=-1.6$ mV.}}
      \label{Fig06}
\end{figure}

At the onset of transport the central system is in an eigenstate, the vacuum state $|3)$, and only the
weak coupling to the leads perturbs the system. As the system remains open with respect to electrons 
and energy the perturbation activates electromagnetic or photon processes. The time-scale needed for
the photon active processes depends thus both on the inherent electrodynamic relaxation time for the
particular states and their coupling to states in the leads. We have confirmed this by repeating the
calculations for Fig.\ \ref{Fig-NeNg} and \ref{Fig-occ} with 4 times larger coupling to the leads 
$g_\mathrm{LR}$ leading to qualitative very similar results on the logarithmic scale, but with all 
features shifted to shorter times. In addition, {seen in Fig.\ \ref{Fig06}}, 
we have started the calculations with one or two
photons initially in the cavity instead of none as in Fig.\ \ref{Fig-NeNg} and \ref{Fig-occ}. In these
cases we also get a significant increase in the mean photon number around the same time as when
we start with no photons in the beginning and get in the end the same steady state. 
{The increase in the mean photon number in the intermediate time regime is always just above
one photon.}

{As expected, the steady state is independent of the initial number of photons in the cavity,
but the relaxation of the system towards the steady state can be different as different
intermediate states participate in the process.}

{For our model with the same parameters as in Fig.\ \ref{Fig-NeNg}-\ref{Fig06} no current will remain
through the system in the steady state, as the electrons entering the system through states in the 
bias window can relax to lower lying states and block further charging and current. This situation can be altered
by changing the plunger gate voltage, $V_g$, or the photon energy.}  

If the transport is started at $V_g=-1.9$ mV without photons in the system the initial charging 
will occur through states in the bias window without photon components, $|15)$ and $|16)$ and at a 
much later time, $t>1$ $\mu$s, photon active transitions will participate in the path to the 
steady state.

In the steady state the occupation of the two spin components of the groundstate represent trapped
charge in the central system as these states are below the lowest band edges in the leads at 1.0 meV,
but the two-electron ground state that also contributes to the steady state is slightly above this edge.  

\begin{figure}[htb]
      \includegraphics[width=0.48\textwidth]{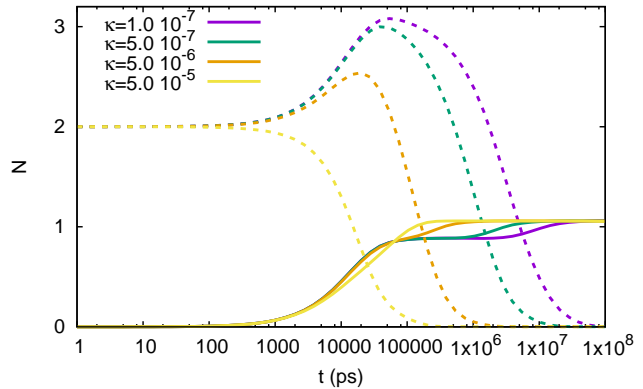}
      \caption{{The mean values of the total electron (bold) and photon (dotted) numbers
               as functions of time for different values of the decay rate, $\kappa$, of the photon cavity.  
               Initially there are 2 photons present in the cavity
               at $t=0$ for $V_g=-1.6$ mV. The unit of $\kappa$ is meV.}}
      \label{Fig07}
\end{figure}

In continuation of the present calculations we have added to the master equation (\ref{GME}) 
a term $+\kappa ([a\rho ,a^\dagger]+[a,\rho a^\dagger])/(2\hbar )$ 
describing Markovian coupling to an external photon reservoir in order to confirm that the photonic 
relaxation channels established here by the coupling to the leads can be blocked by a 
decay rate $\kappa > 10^{-6}$ meV. The number of cavity photons in the central system can be
influenced, both by the coupling to an external photon reservoir, and by the coupling
to external leads acting as electron reservoirs. This can be seen in Fig.\ \ref{Fig07}.
Again, the steady state of the central system is independent of the cavity photon decay rate, $\kappa$,
but the mean electron number in the intermediate time regime depends on it. Still higher decay rates lead
to photon cascades between pure photonic states of the cavity.

In conclusion, we have presented a general mathematical procedure to recast a  
non-Markovian master equation for a complex system of the Nakajima-Zwanzig type in Fock space 
into a Markovian master equation in Liouville space, and subsequently used it to effectively 
explore regimes of dissimilar relaxation channels active, radiative or not, 
at different time-scales in electron transport through a photon cavity.

\begin{acknowledgments}
We acknowledge discussion with Prof.\ Andreas Buchleitner.
This work was financially supported by the Research Fund of the University of Iceland,
and the Icelandic Instruments Fund. We also acknowledge support from the computational 
facilities of the Nordic High Performance Computing (NHPC), and the Nordic network
NANOCONTROL, project No.: P-13053. HSG acknowledges support from MOST, Taiwan, under Grant No.\
103-2112-M-002-003-MY3.
\end{acknowledgments}
%
\bibliographystyle{apsrev4-1}
\providecommand{\othercit}{}
\providecommand{\jr}[1]{#1}
\providecommand{\etal}{~et~al.}

%
%
\end{document}